\documentclass[prl,aps,twocolumn,epsfig,floats,showpacs]{revtex4}
\usepackage{epsfig}
\usepackage{graphicx}
\usepackage{bm}
\usepackage{amsmath}

\begin{document}

\title{The influence of Coulomb on the liquid-gas phase transition and nuclear
multifragmentation }
\author{F.Gulminelli$^{(1)}$,Ph.Chomaz$^{(2)}$,Al. H. Raduta$^{(3)}$, Ad. R. Raduta$%
^{(3)}$}

\affiliation
{(1) LPC Caen (IN2P3-CNRS/Ensicaen et Universit\'{e}),F-14050 Caen C\'{e}dex,France\\
(2) GANIL (DSM-CEA/IN2P3-CNRS),B.P.5027 F-14021 Caen C\'{e}dex,France%
\\
(3) National Institute of Physics and Nuclear Engineering, Bucharest, POB
MG6, Romania}

\begin{abstract}
The liquid-gas phase transition is analyzed from the topologic properties of
the event distribution in the obervables space. A multi-canonical formalism
allows to directly relate the standard phase transition with neutral
particles to the case where the non saturating Coulomb interaction is
present, and to interpret the effect of Coulomb as a deformation of the
probability distributions and a rotation of the order parameter. This
formalism is applied to a statistical multifragmentation model and
consequences for the nuclear multifragmentation phase transitions are drawn.
\end{abstract}


\maketitle

Since the first pioneering multifragmentation experiments~\cite{historique},
the break-up of a hot nucleus above about $3A.MeV$ excitation energy has
been tentatively associated to a transition from liquid to gas.  
A first order phase transition can be unambiguously defined in a finite system
as an abnormal curvature of a thermodynamic potential e.g. a convexity of
the entropy\cite{topology}. When this anomaly affects the energy axis,
the heat capacity becomes negative. The first experimental signatures of
negative heat capacities have been recently reported for atomic clusters 
\cite{haberland,farizon} and nuclei \cite{norvegians,cneg-D-Agostino}%
. The results on $Na$-clusters have been related to the melting of $Na$%
-matter\cite{haberland}. For nuclei the link with bulk properties is
problematic since the long range Coulomb interaction forbids the definition
of a thermodynamic limit. Nuclear matter is a theoretical construction
assuming the proton charge to be zero and one may worry about the connection
between the properties of charged nuclei and the thermodynamics of uncharged
nuclear matter. In this letter, we propose a systematic way to perform such
a connection. 

Let us develop the link between charged and uncharged systems. The
long range Coulomb interaction can be written as $E_{c}=q^{2}V_{c}$ where $q$
represents the proton charge (in electron charge unit $q=1$) while $%
V_{c}=\sum_{p<p^{\prime }}\alpha /\left| \mathbf{r}_{p}-\mathbf{r}%
_{p^{\prime }}\right| $ with $\alpha =e^{2}/4\pi \epsilon _{0}$. The total
energy of the system $E$ can thus be splitted into two terms $%
E_{tot}=E_{n}+E_{c}=E_{n}+q^{2}V_{c}$. If the system is in contact with a
heat bath (canonical ensemble) we can sort the events as a function of both $%
E_{n}$ and $V_{c}$: 
\begin{equation}
p_{\beta }(E_{n},V_{c})=\frac{1}{Z_{\beta }}W(E_{n},V_{c})e^{-\beta
E_{n}-\beta q^{2}V_{c}}.  \label{monocan}
\end{equation}
The canonical total energy distribution 
is an
integration of the generalized distribution (\ref{monocan}): $p_{\beta
}(E_{tot})=\int dV_{c}p_{\beta }(E_{tot}-q^{2}V_{c},V_{c}).$ The
microcanonical entropy, $S_{tot}(E_{tot})=\log W_{tot}(E_{tot})$, is also
directly related to $p_{\beta }(E_{tot})$: $S_{tot}(E_{tot})=\log p_{\beta
}(E_{tot})+\beta E_{tot}+\log Z_{\beta }$. Eq. (\ref{monocan}) naturally
leads to the definition of a generalized entropy $S(E_{n},V_{c})=\log
W(E_{n},V_{c}).$ The standard entropy can be obtained by a  projection
of the density of states on the total energy $E_{tot}=E_{n}+q^{2}V_{c}$ axis 
\[
W_{tot}(E_{tot})=\int dV_{c}W(E_{tot}-q^{2}V_{c},V_{c})
\]
One can also notice that the projection of $W(E_{n},V_{c})$ on the $E_{n}$
axis provides the entropy of the uncharged system 
\[
S_{n}\left( E_{n}\right) =\log W_{n}\left( E_{n}\right) =\log \int
dV_{c}W(E_{n},V_{c})
\]
 
Since we have introduced two observables $(E_{n},V_{c})$, it is natural to
extend the canonical ensemble by introducing two Lagrange multipliers 
$\beta _{n},\beta _{c}$ leading to the distribution of events in 
the two-dimensional observables space 
\begin{equation}
p_{\beta _{n}\beta _{c}}(E_{n},V_{c})=\frac{1}{Z_{\beta _{n}\beta _{c}}}%
W(E_{n},V_{c})e^{-\beta _{n}E_{n}-\beta _{c}q^{2}V_{c}}  \label{multican}
\end{equation}
where $Z_{\beta _{n}\beta _{c}}$ is the partition sum of this
multi-canonical ensemble corresponding to the independent observation of the
two components $E_{n},V_{c}$ of the energy. In the cases $\beta_{n}=\beta
_{c}$ eq.(\ref{multican}) reduces to the usual canonical distribution of
interacting charged particles (\ref{monocan}) while $\beta _{c}=0$
corresponds to systems not affected by the Coulomb force. All intermediate
values $0<\beta _{c}<\beta _{n}$ give rise to interpolating ensembles
between the charged and the uncharged case, in the same way as the gaussian
ensemble~\cite{challa} gives a continuous interpolation between the
microcanonical and the canonical ensemble. The multicanonical ensemble can
also be considered as the canonical ensemble at temperature $\beta _{n}^{-1}$
of particles charged with an effective charge $\bar{q}^{2}/q^{2}=\beta
_{c}/\beta _{n}.$ In fact we can introduce an effective energy as $%
E_{eff}=E_{n}+\bar{q}^{2}V_{c}$. Then the distribution of the effective
energy is a canonical distribution with a Boltzmann factor $e^{-\beta
_{n}E_{eff}}.$ 
Considering $\beta _{c}=0$ (or $\beta
_{c}\rightarrow 0$ faster then the increase of the charge when the volume of
the system is increased), the multicanonical ensemble is a way to study
the thermodynamic limit. 

Let us now focus on the question of (first order) phase transitions by
studying the convex region of the entropy $S\left( E_{c},E_{n}\right) $.
Eq.(\ref{multican}) shows that the curvature matrix of the
entropy and of the probability distribution $\log p_{\beta _{n}\beta
_{c}}\left( E_{c},V_{n}\right) $ are the same for every set of Lagrange
multipliers $\beta _{n},\beta _{c}.$ This implies that the whole
thermodynamics of the uncharged system can be (at least in principle)
completely reconstructed from the only knowledge of the distribution
probability of the charged system at an arbitrary temperature and pressure.
For example, the uncharged system entropy can be deduced within a constant 
from the statistical distribution of the charged system 
by projecting out the Coulomb
energy after a proper Boltzmann weighting 
\begin{equation}
S_{n}(E_{n})=\log \left( \int dV_{c}e^{\beta
_{c}q^{2}V_{c}}p_{\beta _{n}\beta _{c}}(E_{n},V_{c})\right)-\beta _{n}E_{n} .
\label{EQ:reweighting}
\end{equation}
The only limitation can be a practical one for experiments or simulations:
to accumulate enough events at every location.

In order to illustrate these general ideas and make some connections to the
nuclear multifragmentation experiments, we have made some calculations in
the multi-canonical ensemble with a statistical multifragmentation model~%
\cite{raduti}. The model describes a multifragmentation event as an ensemble
of spherical, non overlapping fragments interacting through their mutual
Coulomb repulsion. The masses of the fragments are parametrized from a
charged liquid drop model while their excitation energies are taken from a
Fermi gas distribution with a high energy cut-off. The fragment
translational degrees of freedom are treated classically. 
Each event is characterized by its total mass and charge number ($i.e.$ the
size of the disassembling nuclear source), its total energy that can be
decomposed in a Coulomb $E_{c}$ part that includes also Coulomb effects on
the binding energy, and a non Coulomb one that corresponds to the nuclear
interaction inside fragments plus the translational energy. Each event can
also be associated to a spatial extension, given by the box ('freeze-out'
volume) that contains all the fragments. In an open system as a
multifragmenting nucleus, this volume is not fixed but can
fluctuate from one event to the other. 
If the average volume at
freeze-out is finite, we can treat the volume as an extra observable known
in average taking this constraint into account as an additional Lagrange
multiplier $\lambda =\beta P$ where $\beta =\beta_n $ is the inverse 
temperature and $P$ has the dimension of a pressure ~%
\cite{duflot,raduti_cneg}. 
The multi-canonical probability (\ref{multican}) is then evaluated with a
Metropolis technique \cite{raduti}.

The resulting distribution of events can then be plotted in the $%
(E_{n},V_{c})$ plane\thinspace or equivalently in the $(E=E_{n}+V_{c},V_{c})$
plane. When the Coulomb interaction is at play ($\bar{q}=1$) the total
energy is simply the $E$ axis. When the Coulomb energy is not effective 
($\bar{%
q}=0$) the relevant energy axis is the $E-V_{c}$ one i.e. the second
bisector.

In the isobar ensemble we are considering, energy is expected to be an order
parameter for the ordinary liquid-gas phase transition (in the uncharged
system), implying that, for all pressures below the critical one and for all
temperatures in the transition region, we expect the event cloud to separate
into two components (phases) along the energy axis~\cite{topology}. 
 
\begin{figure}[tbh]
\begin{center}
\end{center}
\caption{ Event distribution in the Coulomb energy versus total energy plane
and relative projections with $\lambda=3\cdot 10^{-4}fm^{-3}$ for a system
of total mass number $A=50$ and atomic number $Z=23$ in the isobar
multicanonical multifragmentation model. Levels of grey and full lines: $%
\beta _{n}^{-1}=3.7MeV$, $\beta _{c}=0$; contour lines and dashed lines: $%
\beta _{n}^{-1}=\beta _{c}^{-1}=3.1MeV$.}
\label{fig:1}
\end{figure}

Figure 1 shows the distribution of events for a system of 50 particles at a
subcritical pressure $\beta P=3\cdot 10^{-4}fm^{-3}$ in the Coulomb energy $%
V_{c}$ and total energy $E$ (calculated respect to the ground state of the 
source) plane. The charged case $\beta _{c}=\beta _{n}$ 
and the uncharged one $\beta _{c}=0$ 
are displayed together with their projections on the two axes.

Let us look at the uncharged case first. The topology of events is
characteristic of a first order phase transition, with an accumulation of
events at low excitation energy and high Coulomb energy (compact
configurations or ''liquid'' phase) and an accumulation at high energy and
low Coulomb (rarefied configurations or ''gas'' phase). The distance between
the two peaks along the nuclear energy axis $E-V_{c}$, measures the latent
heat of the transition while the region of inverted curvature is related to
a negative heat capacity in the microcanonical ensemble. The representation
in the $(E,V_{c})$ plane 
reveals that the liquid peak is constituted by a series of structures. They
can be attributed to different channel openings associated to similar
excitation energies but corresponding to different proton partitions leading
to different $V_{c}$. They constitute the microscopic origin of the global
collective phase change that in the bulk limit
will become the ordinary liquid-gas phase transition. 
 
\begin{figure}[tbh]
\begin{center}
\end{center}
\caption{ Event distribution in the Coulomb energy versus total energy plane
 with $\lambda=5\cdot 10^{-3} fm^{-3}$ for a system
of total mass number $A=50$ and atomic number $Z=23$ in the isobar
multicanonical multifragmentation model. Levels of grey: $%
\beta_n^{-1}=7.5 MeV$, $\beta_c=0$; contour lines in the left panel: 
$\beta_n^{-1}=\beta_c^{-1}=6.7 MeV$; contour lines in the
right panel: distribution with Coulomb reweighted to get the uncharged case.}
\label{fig:2}
\end{figure}

The event distribution of the charged system in Figure 1 shows globally the
same structure as in the uncharged case, with two important differences.
First, to have the same height in the two relative maxima the temperature
must be lower. The decrease of the transition temperature is due to the
repulsive character of the Coulomb interaction~\cite{raduti_cneg}. Second,
even if the phase transition is still clearly visible in the bidimensional
representation, this is not true any more for the projection over the total
energy axis ($E$). Indeed, the fact that the distance between the peaks
(latent heat) is comparable to the width of each peak implies that the
bimodality is hardly perceptible in the energy direction. The usual
interpretation  is then the lowering of the critical point due to the
Coulomb interaction. The complete bidimensional information of Figure 1
demonstrates that this conclusion 
is not completely correct. 
Indeed while the event distribution is almost unchanged, the projection
axis is different in the charged case and the best order parameter (%
$i.e.$ the direction that separates at best the two phases) is now almost
perpendicular to the energy direction. Introducing a better observable
playing the role of the order parameter, $O=aE+bV_{c}$ , would thus restore
the overall picture of the first order phase transition even in the
charged system. 

The introduction of the charge is just a reweighting in the occupation
probabilities of the very same entropy surface. Indeed, if we consider a
charged system at the temperature $\beta _{1}^{-1}$ described by the
distribution $p_{\beta _{1}}^{c}\equiv p_{\beta _{n}=\beta _{1},\beta
_{c}=\beta _{1}}$, the distribution $p_{\beta
_{2}}^{n}\equiv p_{\beta _{n}=\beta _{2},\beta _{c}=0}$ at the temperature $%
\beta _{2}^{-1}$ in the case where the Coulomb interaction is not active
 is given (within a normalization) by 
\begin{equation}
p_{\beta _{2}}^{n}\left( E_{n},V_{c}\right) \propto p_{\beta _{1}}^{c}\left(
E_{n},V_{c}\right) e^{\left( \beta _{1}-\beta _{2}\right) E_{n}+\beta
_{1}q^{2}V_{c}}
\end{equation}

\smallskip An example of the reconstruction of an uncharged  
distribution, starting from a calculation
including the Coulomb interaction is shown in figure 2. The perfect
reproduction of the distribution demonstrates that, even if  the inclusion
of Coulomb leads the system to occupy the phase space in a
different way,  the density of states $W$ itself does not change and the
same information about its convexity anomalies and phase transitions
properties can be 
inferred from the two calculations.

\begin{figure}[tbh]
\begin{center}
\end{center}
\caption{ Event distribution in the Coulomb energy versus total energy plane
with $\lambda =1.26\cdot 10^{-3}fm^{-3}$ for a system of total mass number $%
A=200$ and atomic number $Z=82$ in the isobar multicanonical
multifragmentation model. Levels of grey: $\beta _{n}^{-1}=5.6MeV$, $\beta
_{c}=0$; contour lines (from top to bottom): $\beta _{n}^{-1}=\beta
_{c}^{-1}=3.,3.7,4.4,4.8MeV$. Black contour lines: $\beta _{n}^{-1}=5.MeV$
and partially screened charge $\bar{q}=0.14$. }
\label{fig:3}
\end{figure}

From a practical point of view, this reconstruction is possible only for the
energy states where the population sample is statistically significant; if
the probability distributions are concentrated in very different locations
in the observables space the size of the sample needed will become
increasingly (and soon prohibitively) high. This more complicated situation
happens when the average partitions are very different in the charged and
uncharged case. This is indeed the case if we consider heavily charged
systems, for which the Coulomb distortions get more appreciable. Figure 3
shows the event distribution for a heavy system composed of $A=200$
particles. When the Coulomb interaction is neglected the situation gets
closer to the expected behavior in the bulk: the channel openings are not
recognizeable any more and only the collective liquid-like to gas-like state
change survives, with a minimum between the two phases getting deeper. If
Coulomb is switched on, the event topology drastically changes. A much
smaller portion of phase space is populated by any calculation at a single
temperature, and this stays true if the pressure is changed. The event
distribution 
turns around the first order phase transition without diving into the
coexistence region. The external edge of the anomalous curvature region is
only touched by the calculations at the lowest temperature, pointing to an
order parameter almost perfectly perpendicular to the energy direction. This
residual bimodality corresponds to the opening of asymmetric fission, and is
the only rest of the liquid-gas phase transition in heavily charged systems.
Since the average transformations in the $(E,V_{c})$ plane are so far away,
the reconstruction of the neutral matter phase diagram requires a
prohibitively high statistics and cannot be performed for too heavy systems. 

In conclusion, in this letter we have introduced a multi-canonical formalism
that allows a direct mapping between a system composed of particles with or
without an electric charge. This general framework allows to study the
effect of a long range non saturating force as the Coulomb interaction on a
first order phase transition. Such an interaction is seen to deform the
distribution of events in the space of observables but its strongest effect
is to rotate the energy axis away from the order parameter. This effect is
responsible for the apparent reduction or suppression of the coexistence
region. If the system is not too heavily charged, the nuclear energy 
remains well correlated with the order parameter
and the transition
phenomenology is not strongly affected by the presence of the Coulomb field. 
 The Coulomb simply manifests itself as a lowering of the critical parameters
which is normal since the binding energy is reduced. 
The convexity anomalies of the entropy can be traced back from the event
distribution independently of the Coulomb interaction and the equations of
state of the uncharged system can be obtained from a charged population
sample if the Coulomb energy is measured on an event-by-event basis and the
relevant phase space is sampled with enough statistics.

For heavily charged systems the
considered statistical model predicts that, at low temperature, the events
strongly differ from the uncharged partitions. 
Therefore the two thermodynamics can only be
related at high energy when the gas phase is reached. 
Moreover in the charged system the transition from the liquid to gas
goes around the coexistence region and so corresponds to a cross over and
not to a first order phase transition. However, the size at which this
phenomenon appears is model dependent and more investigations are needed
to get quantitative predictions.

Finally we would like to stress that all the interpolating ensembles
with  $0<\beta_{c}<\beta _{n}$ (see black contour lines in figure \ref{fig:3})
may also be interesting to model experimental situations in which
the different parts of the energy are not equilibrated. This may happen in
actual reactions since the relaxation time depends critically on
the range of the force~\cite{latora}. For example, many scenarii of
multifragmentation involve spinodal instabilities which are weakly sensitive
to the Coulomb force \cite{spinodal}. 
Being very chaotic, spinodal decomposition may then
well populate a complete phase space for which $E_{n}$ is the
important physical quantity and not $E_{tot}$ . This would
correspond to a $\beta _{c} \approx 0$ case. 
Such a situation might be suggested by
the insensitivity to the total system charge 
of the fragmentation pattern observed in
multifragmentation data from central reactions and predicted in spinodal
decomposition simulations\cite{scaling}. However, this question requires a
strong theoretical and experimental effort which goes beyond the scope of
the present article.

\end{document}